\documentstyle[12pt]{article}
\begin{document}
\vspace{1.cm}
\begin{center}
\    \par
\    \par
\    \par
\     \par

              {\bf{ LIGHT-FRONT ANALYSIS OF $\pi^{-}$ MESONS PRODUCED
           IN \\Mg - Mg COLLISIONS AT  4.3 A GeV/c.}}
\end{center}
\   \par
\    \par
\par
    {\bf{ M. Anikina$^{1}$, L.Chkhaidze$^{2}$, T.Djobava$^{2}$,
     V.Garsevanishvili $^{3,4}$, \par
L.Kharkhelauri$^{2}$}}\par
\  \par
$^{1}$ Joint Institute for Nuclear Research, 141980 Dubna, Russia\par
$^{2}$ High Energy Physics Institute, Tbilisi State University,\par
\hspace{.1cm} University Str. 9, 380086 Tbilisi, Republic of Georgia\par
$^{3}$ Mathematical Institute of the Georgian Academy of Sciences \par
\hspace{.1cm} M.Alexidze Str. 1 , 380093 Tbilisi, Republic of Georgia  \par
$^{4}$ CERN, CH-1211, Geneva 23, Switzerland\par
\   \par
E-mail: djobava@sun20.hepi.edu.ge     \par
\begin{center}

                     \bf{ ABSTRACT }
\end{center}
\par
Light-front analysis of $\pi^{-}$ mesons in
 Mg-Mg
 collisions is carried out. The phase space of secondary
pions is naturally divided into two parts in one of which the thermal
equilibration assumption seems to be in a good agreement with data. Corresponding
temperatures are extracted and compared to the results of other
experiments.
The experimental results have been compared with the predictions of the Quark
Gluon
String Model (QGSM) and satisfactory  agreement between
 the experimental data and the model has been found.
\par
\hspace{2cm} measured pion distributions; the analysis in light front variables\par
\hspace{2cm} is carried out; deduced thermal equilibration, which is characterized\par
\hspace{2cm} by the temperature T; Comparison with Quark Gluon String Model.
\pagebreak
\begin{center}
\bf { 1.  INTRODUCTION }
\end{center}
\  \par
\par
In the experiments with beams of relativistic heavy ions one hopes
to observe the extreme conditions when the phase transitions in nuclear
matter are expected (see, e.g. [1-3]).
\par
   For the experimental study of such  transitions it is necessary to
understand the  mechanism  of  collisions  and investigate
the characteristics of multiparticle production  in nucleus-nucleus
interactions. The study of single particle inclusive processes is
one of the simplest and effective tools for the understanding of
dynamics of multiple production of secondaries.
\par
   In this  respect  it  is important  to  investigate  the
properties of $\pi^{-}$ mesons, which are predominantly produced particles
carrying the information about the dynamics of collision and which are
reliably  identified.  Besides,  the  pion   production   is   the
predominant process at Dubna energies.
\par
   Our previous results on pion production experiment (cross-sections,
multiplicities, rapidities, transverse momenta, intercorrelations
between various characteristics, etc) using  the
streamer chamber spectrometer  SKM-200 and its modified version GIBS
   in  inelastic  and  central
nucleus-nucleus interactions are presented in [4-6].
\par
In this paper we present the light front analysis of $\pi^{-}$ mesons
produced in Mg-Mg collisions. It is assumed that interactions of identical
nuclei
( Mg-Mg) give the possibility of better manifestation of nuclear
effects than the interactions of asymmetric pairs.
In some cases light front analysis [7] seems to be more sensitive
to the details of
the interaction mechanism as compared to the presentation of data in
terms of the well known Feyman variables $x_{F}$, rapidity  $Y$ etc.
\  \par
\  \par
\begin{center}
\bf { 2.  EXPERIMENT }
\end{center}
\  \par
\par
    The data were obtained using the  SKM-200  facility
and its modified version GIBS of the Dubna
Joint Institute for Nuclear Research.  SKM-200-GIBS   consists  of  a   2m
streamer chamber, placed in a magnetic field of  $\sim$ 0.8  T  and   a
triggering system. The streamer chamber was exposed by beam  of
 Mg  nuclei accelerated in the synchrophasotron  up
to a momentum of  4.3 GeV/c  per incident nucleon . The solid target (Mg)
in the form of thin disc with thickness 1.56 g/cm$^{2}$
was mounted within the fiducial
volume
of the chamber. The triggering system  allowed  the  selection  of
"inelastic"  and "central" collisions.
\par
   The inelastic trigger selected all inelastic  interactions of
the incident nuclei on the target.
 The central trigger selected  events defined as those
without charged  and neutral projectile spectator fragments
 ( P/Z $>$ 3 GeV/c) within a cone of half angle
 $\Theta_{ch}$=$\Theta_{n}$=$2.4^{0}$
 (the trigger efficiency was
99$\%$ for events  with  a single charged particle
and 80$\%$  for events with a single neutron in the cone).
The  trigger  mode  for each exposure is defined as \\ T ($\Theta_{ch}$
,$\Theta_{n}$ ) ($\Theta_{ch}$  and  $\Theta_{n}$  expressed  in degrees and rounded
to the closest  integer  value). Thus Mg-Mg interactions obtained on
the set-up correspond to the trigger T(2,2). The fraction of such
events was $\approx~ 4\cdot10^{-4}$ among all inelastic interactions.
The experimental setup and the logic of the triggering systems are
presented in Fig.1.
\par
  Primary results of scanning and measurements were biased due to several
experimental effects and appropriate corrections were introduced. The biases
and correction procedures were discussed in detail in [4,6].
  Average measurement errors of the  momentum  and
production angle  determination for $\pi^{-}$ mesons are
$<\Delta P/P>$= 1.5$\%$, $\Delta$$\Theta$ =0.1$^{0}$.
\  \par
\  \par
\  \par
\begin{center}

     \bf{ 3. LIGHT FRONT PRESENTATION OF INCLUSIVE DISTRIBUTIONS}
\end{center}
\   \par
\par
An important role in establishing of many properties of multiple production
is played by the choice of kinematical variables in terms of which observable
quantities are presented (see in this connection, e.g. [8]).

Here we propose  unified scale invariant variables for the presentation of single
particle inclusive distributions, the properties of which are described
below.

Consider an arbitrary 4--momentum $p^{}_{\mu}(p^{}_0,\vec p)$ and introduce the light
front combinations [9]:
\begin{eqnarray}
p^{}_{\pm}=p^{}_0\pm p^{}_3 \label{eq1}
\end{eqnarray}

If the 4--momentum $p^{}_{\mu}$ is on the mass shell $(p^2=m^2)$, the combinations
$p^{}_{\pm},\ \vec p^{}_T$ (where $\vec p^{}_T=(p^{}_1,\ p^{}_2)$) define the so called
horospherical coordinate system (see, e.g. [10]) on the
corresponding mass shell hyperboloid
$p^2_0-\vec p\, ^2=m^2$.

Let us construct the scale invariant variables [7]:
\begin{eqnarray}
\xi^{\pm}=\pm {p^c_{\pm}\over{p^a_{\pm}+p^b_{\pm}}} \label{eq2}
\end{eqnarray}

\noindent in terms of the 4--momenta $p^a_{\mu},\ p^b_{\mu},\ p^c_{\mu}$ of particles
$a,\ b,\ c$, entering the inclusive reaction $a+b\to c+X$.
The $z$-axis is taken to be the collision axis, i.e. $p_{z}=p_{3}=p_{L}$.
 Particles $a$
and $b$ can be hadrons, heavy ions, leptons.
The light front variables $\xi^{\pm}$ in the centre of mass frame
are defined as follows [7]:
\begin{eqnarray}
\xi^{\pm}&=&\pm {E\pm p_z\over{\sqrt{s}}}=\pm {E+|p_z|\over{
\sqrt{s}}}  \label{eq3}
\end{eqnarray}
where $s$ is the usual Mandelstam variable,
$E=\sqrt{p^{2}_z+p^{2}_T+m^{2}}$ and $p_{z}$ are
the energy and the $z$ - component of the momentum of produced particle.
The upper sign in Eq. (3) is used for the right hand side hemisphere and
the lower sign for the left hand side hemisphere. It is convenient also to
introduce the variables
\begin{eqnarray}
\zeta^{\pm}=\mp{\rm ln}|\xi^{\pm}| \nonumber
\end{eqnarray}
in order to enlarge the scale in the region of small $\xi^{\pm}$.

The invariant differential cross section in terms of these variables
looks as follows:
\begin{eqnarray}
E{d\sigma\over{d\vec p}}={|\xi^{\pm}|\over{\pi}}\ {d\sigma\over{
d\xi^{\pm}dp^{2}_T}} = {1\over{\pi}}\ {d\sigma\over{
d\zeta^{\pm}dp^{2}_T}}  \label{eq4}
\end{eqnarray}
In the limits of high $p_{z}$ ($|p_z|\gg p_T$)
and high $p_{T}$  ($p_T\gg |p_z|$) the $\xi^{\pm}$
variables go over to the well known variables
$x_{F}={2p_{z}}/{\sqrt{s}}$ and
$x_{T}= {2p_{T}}/{\sqrt{s}}$, respectively, which are intensively
used in high energy physics.
$\xi^{\pm}$--variables are related to $x^{}_F$,\ $x^{}_T$ and rapidity
$y$ as follows:
\begin{eqnarray}
\xi^{\pm}&=&{1\over{2}}\left(x^{}_F\pm \sqrt{x^2_F+x^2_{T}}\right)\
;\ x^{}_{T}={2m_T\over{\sqrt{s}}} \label{eq5}\\
y&=&\pm {1\over{2}}\ {\rm ln}\, {(\xi^{\pm}\sqrt{s})^2\over{m^{2}_T}}
~;~ m_{T}=\sqrt{p^{2}_{T}+m^{2}}
\label{eq6}
\end{eqnarray}
\par
The principal differences of $\xi^{\pm}$ distributions as compared to the
corresponding $x_{F}$ -- distributions are the following:
1) existence of some forbidden region around the point $\xi^{\pm}=0$,
2) existence of maxima at some $\tilde{\xi^{\pm}}$ in the region of
 relatively small $|\xi^{\pm}|$, 3) existence of the limits for
$\vert\xi^{\pm}\vert \leq m/\sqrt{s}$.
 The maximum
at $\tilde{\zeta}^{\pm}$ is also observed in the invariant differential cross
section
$\displaystyle {1\over{\pi}}\ {d\sigma\over{d\zeta^{\pm}}}$. However, the region
$|\xi^{\pm}|>|\tilde{\xi}^{\pm}|$ goes over to the region
$|\zeta^{\pm}|<|\tilde{\zeta}^{\pm}|$ and vice
versa.
\par
Note that the light front variables have been introduced long time ago by
Dirac [9] and they are widely used now in the treatment of
many theoretical problems
(see, e.g. original and review papers [11-19] and references therein). They have
been used also in a number of phenomenological applications (see, e.g. [20]).
\  \par
\  \par
\  \par
\begin{center}
\bf{4. THE ANALYSIS OF PION DISTRIBUTIONS IN TERMS OF LIGHT FRONT
VARIABLES.}
\end{center}
\  \par
\  \par
\par
The analysis has been carried out for the $\pi^{-}$ mesons from
central (trigger T(2,2)) Mg-Mg collisions
($\sim$ 6200 events, $\sim$ 50 000 $\pi^{-}$
mesons).
In Figs.2 and 3 $x_{F}$ -- and $\xi^{\pm}$ -- distributions of all
$\pi^{-}$ mesons are presented.
The experimental data
for invariant distribution $(1/\pi) \cdot dN/d\zeta^{\pm}$
are shown in Fig.4. The curve is the result of
the polinomial
approximation of the experimental distribution and the
maximum is observed  at $\tilde{\zeta^{\pm}}=2.0 \pm 0.3$.
The value $\tilde{\zeta^{\pm}}$
is the boundary of the two regions with significantly
different characteristics of secondaries.
\par
In Figs.5 and 6 the $p_{T}^{2}$ and the angular distributions
 of $\pi^{-}$ mesons in the forward hemisphere
 in different regions of $\zeta^{+}$ ( $\zeta^{+} >
\tilde{\zeta^{+}}$
and $\zeta^{+} < \tilde{\zeta^{+}}$) are presented.
Similar results have been
obtained for the backward emitting $\pi^{-}$ mesons.
\par
One can see from Figs.5 and 6, that the $p_{T}^{2}$
and angular
 distributions of $\pi^{-}$ mesons differ significantly in
$\zeta^{+} > \tilde{\zeta^{+}}$ ($\xi^{+} < \tilde{\xi^{+}}$)
and $\zeta^{+} < \tilde{\zeta^{+}}$ ($\xi^{+} > \tilde{\xi^{+}}$)
regions. The angular distribution of pions in the region $\zeta^{+} <
\tilde{\zeta^{+}}$
is sharply anisotropic in contrast to the almost flat distribution
in the region $\zeta^{+} > \tilde{\zeta^{+}}$. The slopes of
$p_{T}^{2}$ -- distributions differ greatly in different regions of
$\zeta^{+}$.
The average values $<p_{T}^{2}>$ in these two regions also differ:
$<p_{T}^{2}>=(0.027\pm0.002$) (GeV/c)$^2$ in the region  $\zeta^{+} > \tilde{\zeta^{+}}$;
$<p_{T}^{2}>=(0.103\pm0.009$)  (GeV/c)$^2$ in
      $\zeta^{+} < \tilde{\zeta^{+}}$.
\par
The flat behaviour of the angular distribution allows one to think that
one observes a partial thermal equilibration in the region
$|\xi^{\pm}| < |\tilde{\xi^{\pm}}|$ ( $|\zeta^{\pm}| > |\tilde{\zeta^{\pm}}|$)
of phase space.
\par
Note, that the paraboloids of constant $\xi^+$
\begin{eqnarray}
p_z={p^{2}_T+m^{2}-(\tilde{\xi}^+\sqrt{s})^2\over{-2\tilde{\xi}^+
\sqrt{s}}} \label{eq7}
\end{eqnarray}

\noindent separates two groups of
particles with significantly different characteristics.
\par
To describe the spectra in the region $\zeta^{+} > \tilde{\zeta^{+}}$ the
Boltzmann ~~~
$  f(E)\sim e^{-E/T} $
~~ and the Bose-Einstein (B-E) ~~~
$  f(E)\sim (e^{-E/T} - 1)^{-1}$
~~ distributions have been used.
\par
 The distributions
($1/\pi) \cdot  dN/d\zeta^{+}$, $dN/dp_{T}^{2}$, $dN/dcos\Theta$ look in
this region as follows~:
\begin{eqnarray}
{d\sigma\over{dp^2_T}}&\sim&\int_0^{p^{}_{z,max}}f(E)dp^{}_z \label{eq8}\\
{d\sigma\over{d\cos\theta}}&\sim&\int_0^{p^{}_{max}}f(E)p^2dp \label{eq9}\\
{1\over{\pi}}\ {d\sigma\over{d\zeta^+}}&\sim&\int_0^{p^2_{T,max}}Ef(E)dp^2_T
\label{eq10}
\end{eqnarray}
where:
\begin{center}
${p^{2}_{T,max}} =(\tilde{\xi^{+}}\sqrt{s})^{2} - m_{\pi}^{2}$
\end{center}
\begin{center}
${p_{z,max}} =[p_{T}^{2}+m^{2}-(\tilde{\xi^{+}}\sqrt{s})^{2}]/
(-2\tilde{\xi^{+}}\sqrt{s})$
\end{center}
\begin{center}
$p_{max}=(-\tilde{\xi^{+}}\sqrt{s}cos\Theta + \sqrt{(
\tilde{\xi^{+}}\sqrt{s})^{2}- m_{\pi}^{2}
sin^{2}\Theta})/sin^{2}\Theta$
\end{center}
\par
The experimental distributions in the region $\zeta^{+} >\tilde{\zeta^{+}}$
have been fitted by the expressions (8), (9), (10), respectively. The results
of the fit are given in Table 1 and Figs. 5 $\div$ 7
(by the Boltzmann distribution).
They show a rather good agreement with experiment. One can see from the Table
1, that the values of the parameter $T$ extracted by fitting the data
with Boltzmann and Bose-Einstein distributions coincide within the errors.
\par
In the region $\zeta^{+} < \tilde{\zeta^{+}}$ the $p_{T}^{2}$ -- distributions
has been fitted by the formula
\begin{eqnarray}
\frac {dN}{dp_{T}^{2}} \sim \alpha \cdot e^{-\beta_{1}P_{T}^{2}} +
(1-\alpha) \cdot e^{-\beta_{2}p_{T}^{2}} \label{eq11}
\end{eqnarray}
and the $\zeta^{+}$ -- distributions by the formula
\begin{eqnarray}
\frac{1}{\pi}\cdot\frac {dN}{d\zeta^{+}} \sim (1 - \xi^{+})^{n}=
(1 - e^{-\vert \zeta^{+}\vert})^{n} \label{eq12}
\end{eqnarray}
 The results of the fit are given in
Table 2 and Figs.5 and 7.
\par
Thus the spectra of $\pi^{-}$ mesons in the region $\zeta^{+} >
\tilde{\zeta^{+}}$ are
satisfactorily described by the formulae which follow from
the thermal equilibration. The same formulae when extrapolated to the region
$\zeta^{+} < \tilde{\zeta^{+}}$ (Fig.7) deviate significantly from the data.
On the other hand the dependence $(1 - \xi^{+})^{n}$ is in a good
agreement with experiment in the region $\zeta^{+} < \tilde{\zeta^{+}}$ and
deviates from it in the region $\zeta^{+} > \tilde{\zeta^{+}}$ (Fig.7).
\par
Several theoretical models of nucleus-nucleus collisions at
high energy have been
proposed [21]. In this paper Quark Gluon String Model
(QGSM) [22] is used for a comparison with experimental data. The QGSM is based
on the Regge and string phenomenology of particle production in inelastic
binary hadron collisions [23].
The QGSM simplifies the nuclear effects (neglects the potential
interactions between hadrons, coalescence of nucleons and etc.).
A detailed description and
comparison of the QGSM with experimental data in a wide energy range can be
found in paper [24].
 To describe the evolution of the hadron and
quark-gluon phases, this model uses a coupled system of Boltzmann-like kinetic
equations. Nuclear collisions are treated as a mixture
of independent interactions of the projectile and target nucleons, stable
hadrons and short lived resonances. QGSM includes low mass vector mesons
and baryons with spin 3/2, mostly  $\Delta$(3/2,3/2)  resonances.
The procedure of generation consists of 3 steps: the definition of
configuration of colliding nuclei, production of quark-gluon strings
and fragmentation of strings (breakup) into observed hadrons. The formation
time of hadrons is also included in the model. The QGSM has been extrapolated
to the range of intermediate energy (${\sqrt{s}}$ $\leq$ 4 GeV)
 to use it as a basic process
during the generation of hadron-hadron collisions.
After
hadronization the newly formed secondary hadrons are allowed to rescatter.
To determine the interaction between hadrons, the experimental total, elastic,
and annihilation cross sections have been used.
Isotopic invariance and  the additive quark model relations  were used
to avoid data deficiency. The resonance  cross sections were assumed to be
identical to the
 stable particle cross sections with the same quark
content.
 At low energy
the QGSM reduces to a standard cascade model without mean
field effects.
 The model
yields a generally good overall fit to most experimental data [24,25].
Particularly the model describes well the average kinamatical characteristics
and distributions of pions in Mg-Mg interactions [26].
\par
We have generated Mg-Mg interactions using Monte-Carlo generator
COLLI, based on the QGSM. The events had been traced through the detector
and trigger filter.
\par
In the generator COLLI there are two possibilities to generate events:
1) at not fixed impact parameter $\tilde{b}$ and 2) at fixed $b$.
The events have been generated for $\tilde{b}$
(0$\leq \tilde{b} \leq$6  fm).
  The number of simulated events is $\sim$ 4000.
   From the impact parameter distribution
 we obtained the mean value of
$<b>=1.34\pm0.02$ fm.
For the obtained value of $<b>$, we have generated
a total sample of 6200 events.
The two regimes
are consistent and it seems, that in our experiment the value of
b=1.34 fm for Mg-Mg is most probable.
\par
The experimental results have been compared with the predictions of the QGSM
 for b=1.34 fm and satisfactory  agreement between
 the experimental data and the model have been found.
In Figs 2, 3 and 4 the $x_{F}$, $\xi^{\pm}$ and $\zeta^{\pm}$ -
distributions
of $\pi^{-}$ mesons from the QGSM calculations are presented
together with the experimental ones. One can see, that the QGSM reproduces
these distributions.
The QGSM also reproduces the $ p_{T}^{2} $  and $ cos\Theta $
distributions (Figs.5 and 6). The QGSM data show the similar characteristics
in the different regions of $\zeta$ as experimental ones:
 sharply anisotropic angular distributions in the region $\zeta^{+} <
\tilde{\zeta^{+}}$  and the almost flat distribution
in the region $\zeta^{+} > \tilde{\zeta^{+}}$. The slopes of
$p_{T}^{2}$ -- distributions differ greatly in different regions of
$\zeta^{+}$.
The average values $<p_{T}^{2}>$ in these two regions also differ:
$<p_{T}^{2}>=(0.029\pm0.003$) (GeV/c)$^2$ in the region  $\zeta^{+} > \tilde{\zeta^{+}}$;
$<p_{T}^{2}>=(0.109\pm0.009$)  (GeV/c)$^2$ in
      $\zeta^{+} < \tilde{\zeta^{+}}$. The average values of $<p_{T}^{2}>$
in the different regions of $\zeta$ from the experimental and QGSM data
are in a good agreement between each other.
The distributions obtained by the
 QGSM in the region $\zeta^{+} >\tilde{\zeta^{+}}$
have been fitted by the expressions (8), (9), (10). The results
of the fit are given in Table 1 and Figs. 5 $\div$ 7.
In the region $\zeta^{+} < \tilde{\zeta^{+}}$ the $p_{T}^{2}$
and the $\zeta^{+}$ -- distributions
have been fitted by the formulae (11) and (12), respectively.
 The results of the fit are given in
Table 2 and Figs.5 and 7(by the Boltzmann distribution).
One can see from the Table 1, that
the values of the $T$ extracted from the experimental and QGSM data
coincide within the errors, only for the $ (1/\pi) \cdot dN/d\zeta^{+}  $ distribution
the QGSM slightly overestimates the experimental result (for the B-E
distribution).
\par
In our previous article [27] the $\pi^{-}$ mesons temperature
in Mg-Mg collisions was estimated by means of inclusive kinetic
energy and transverse momentum spectra for $\pi^{-}$ mesons: in
rapidity interval 0.5~$\div$~ 2.1, which corresponds to the pionization
region for $\pi^{-}$ mesons and
with the c.m.s. angles $90^{o} \pm 10^{o}$.
The pion spectra have been fitted by a sum of two exponentials,
or two temperatures $T_{1}$ and $T_{2}$. $T_{1}$ = 55$\pm$1 MeV,
$T_{2}$=113$\pm$2 MeV. This explained by two mechanisms
of pion production: directly ($T_{2}$) and via $\Delta$ resonance decay ($T_{1}$).
The relative yield of $T_{2}$ is $\approx 22 \%$.
One can see, that the light front analysis gives
temperature which is weighted average
value of $T_{1}$ and $T_{2}$.
The temperatures of pions have been extracted
in the GSI experiments (FOPI, FRS,
KAON and TAPS-
Collaborations, see, e.g. [28,29]) similarly.
At FOPI Collaboration [28] it has been obtained that the $\pi^{-}$
spectra from Ni-Ni collisions
require the sum of two exponential functions with independent
yields and slope parameters $T_{l}$ and $T_{h}$ describing mostly the
low and the high momentum part of the spectrum, respectively:
at E=1.06 A GeV $T_{l}=55\pm3$ MeV, $T_{h}=93\pm5$ MeV;
at E=1.45 A GeV $T_{l}=56\pm3$ MeV, $T_{h}=100\pm5$ MeV;
at E=1.93 A GeV $T_{l}=61\pm3$ MeV, $T_{h}=115\pm6$ MeV.
 At FRS Collaboration it has been
obtained that the $T$ for $\pi^{-}$ mesons in Ne-NaF collisions range
from 78$\pm$2 MeV to 96$\pm$3 MeV for projectile energies from 1.34 to
1.94 A GeV. At TAPS Collaboration for $\pi^{0}$ mesons:
$T=83\pm3$ MeV
in C-C interactions at incident energy  of E=2 A GeV;
$T=70\pm1$ MeV
in Ar-Ca interactions at incident energy  of E=1.5 A GeV;
At KAON Collaboration the value of $T$ for $\pi^{+}$ mesons
ranges from
 71$\pm$3 MeV (at energy E=1 A GeV) to 95$\pm$3 MeV
(at energy E=1.8 A GeV).
\par
It should be noted that the
extraction procedures of $T$ in the light front variables
 and in the GSI experiments
are different. It seems to be interesting in this connection
to perform the light front analysis of the GSI data.
\   \par
\  \par
\begin{center}
\bf { 5.  CONCLUSIONS }
\end{center}
\  \par
\par
The analysis of $\pi^{-}$ -- mesons from Mg-Mg
collisions in the light front variables has been carried out.
In some region of phase space of $\pi^{-}$
mesons the thermal equilibration seems to be reached,
 which is characterized by the temperature $T= 75 \pm 3$ MeV.
The variables used can serve as a possible convenient tool to study
hadron-production processes in hadron-hadron,
nucleus-nucleus  and $e^{+}~e^{-}$ -- interactions.
\par
A remark on the nature of maxima in $\zeta^{\pm}$ -distributions is in order.
Recently ALEPH Collaboration observed the maxima in the $\xi$ - distributions
($\xi=-ln~ p/p_{max}$) [30] of secondary hadrons in $e^{+}~e^{-}$
collisions, which coincide to high precision with the predictions of
perturbative QCD (see., e.g. [31,32]). The accuracy of coincidence increases
when the next to leading order corrections are taken into account. So the
shapes of $\xi$ - distributions are related to the details of the underlying
dynamics. Similarly, it seems that the maxima in $\zeta^{\pm}$ -distributions
reflect the dynamics of the processes considered.
\   \par
\   \par
\   \par
\par
The authors express their deep gratitude to J.-P.Alard,
N.Amaglobeli, D.Ferenc, Sh.Esakia,
S.Khorozov, G.Kuratashvili, J.Lukstins, J.-F.Mathiot,
Z.Menteshashvili, G.Paic, G.Roche
for interesting discussions.
We are very grateful to N.Amelin for providing us with the QGSM
code program COLLI.
One of the authors (V.G.) would like to thank
A.De Rujula and G.Veneziano and CERN Theory Division for the warm hospitality.
\pagebreak

\newpage
\begin{center}
\bf{FIGURE CAPTIONS}
\end{center}
\  \par
{\bf{Fig.1}}
Experimental set-up. The trigger and the trigger distances are not to
scale. \par
\  \par
{\bf{Fig.2}}
The  $x_{F}$ -- distribution of  $\pi^{-}$ mesons.
$\circ$   --  experimental data,
$\ast$ -- QGSM data.
\par
\   \par
{\bf{Fig.3}}
The  $\xi^{\pm}$ -- distribution of  $\pi^{-}$ mesons.
$\circ$   --  experimental data,
$\ast$ -- QGSM data.\par
The curve is a result of polinomial approximation
of the experimental data.
\par
\   \par
{\bf{Fig.4}}
The  $ \zeta^{\pm} $ distribution of  $\pi^{-}$ mesons.
$\circ$   --  experimental data,
$\ast$ -- QGSM data. The curve -- result of polinomial approximation
of the experimental data.
\par
\  \par
{\bf{Fig.5}}
The $ p_{T}^{2} $ distribution of  $\pi^{-}$ mesons.
$\circ$   -- experimental data
for $\zeta^{+} > \tilde{\zeta^{+}}$   ($\tilde{\zeta^{+}}$=2.0),
$\diamond$   -- the QGSM data
for $\zeta^{+} > \tilde{\zeta^{+}}$,
$\bigtriangleup$ -- experimental data for $\zeta^{+} < \tilde{\zeta^{+}}$,
\mbox{\put(3.,	0.){\framebox(6.,6.)[cc]{}}}~~~~ -- the QGSM data
for $\zeta^{+} < \tilde{\zeta^{+}}$. The dashed lines - fit of the
experimental data by the Boltzmann distribution
; the solid lines - fit of the QGSM data by the Boltzmann distribution.
\par
\   \par
{\bf{Fig.6}}
The  $ cos\Theta $ distribution of  $\pi^{-}$ mesons.
$\circ$   -- experimental data
for $\zeta^{+} > \tilde{\zeta^{+}}$   ($\tilde{\zeta^{+}}$=2.0),
$\diamond$   -- the QGSM data
for $\zeta^{+} > \tilde{\zeta^{+}}$,
$\bigtriangleup$ -- experimental data for $\zeta^{+} < \tilde{\zeta^{+}}$,
\mbox{\put(3.,	0.){\framebox(6.,6.)[cc]{}}}~~~~ -- the QGSM data
for $\zeta^{+} < \tilde{\zeta^{+}}$.
 The dashed lines - fit of the
experimental data by the Boltzmann distribution; the solid lines - fit of the
QGSM data by the Boltzmann distribution.
\par
\   \par
{\bf{Fig.7}}
The  $ (1/\pi) \cdot dN/d\zeta^{+}  $ distribution of  $\pi^{-}$ mesons.
$\circ$ -- experimental data, the solid line -- fit of the data in the region
$\zeta^{+} > \tilde{\zeta^{+}}$
by the Boltzman distribution, the dashed line --
fit of the data in the region $\zeta^{+} < \tilde{\zeta^{+}}$  by the formula
$(1 - e^{-\vert \zeta^{+}\vert})^{n}$;
$\bigtriangleup$ -- the QGSM data.
\   \par
\   \par
\begin{center}
\bf{TABLE CAPTIONS}
\end{center}
\   \par
{\bf{Table 1.}}
The
results of the fit of the distributions of $\pi^{-}$ --
mesons in the region  $\zeta^{+} >\tilde{\zeta^{+}}$.\par
\  \par
\  \par
{\bf{Table 2.}} The results of the fit of the  distributions of $\pi^{-}$
mesons in the region  $\zeta^{+} < \tilde{\zeta^{+}}$ .
\newpage
{\bf{Table 1.}}
The
results of the fit of the distributions of $\pi^{-}$ --
mesons in the region  $\zeta^{+} >\tilde{\zeta^{+}}$.\\
\  \par
\  \par
\begin{tabular}{|c|c|c|c|c|c|c|c|}    \hline
 \multicolumn{1}{|c|}{}& \multicolumn{1}{c|}{}&
\multicolumn{6}{c|}{}\\
\multicolumn{1}{|c|}{$ Type $}&
\multicolumn{1}{c|}{$ Number $}&
\multicolumn{6}{c|}{{$T~(MeV)$}}\\
\cline{3-8}
 \multicolumn{1}{|c|}{$of$}&
\multicolumn{1}{c|}{$of$}&
\multicolumn{2}{c|}{}&\multicolumn{2}{c|}{}&\multicolumn{2}{c|}{}\\
\multicolumn{1}{|c|}{$ events$}&
\multicolumn{1}{c|}{ $ events$}&
\multicolumn{2}{c|}{$dN/dP_{T}^{2}$}&
\multicolumn{2}{c|}{$dN/dcos\Theta $}&
\multicolumn{2}{c|}{$1/\pi\cdot dN/d\zeta ^{+}$}\\
\multicolumn{1}{|c|}{}& \multicolumn{1}{c|}{}&
\multicolumn{2}{c|}{}&\multicolumn{2}{c|}{}&\multicolumn{2}{c|}{}\\
\cline{3-8}
\multicolumn{1}{|c|}{}&
\multicolumn{1}{c|}{}&
\multicolumn{1}{c|}{$Boltz$}&
\multicolumn{1}{c|}{$B-E$}&
\multicolumn{1}{c|}{$Boltz$}&
\multicolumn{1}{c|}{$B-E$}&
\multicolumn{1}{c|}{$Boltz$}&
\multicolumn{1}{c|}{$B-E$}\\
\hline
 &&&&&&&\\
   exp. & 6239
& 76 $\pm$ 2 &  76 $\pm$ 3
   & 75 $\pm$ 3 & 78 $\pm$ 4 &  75 $\pm$ 3 & 75 $\pm$ 4\\
 &&&&&&&\\
\hline
 &&&&&&&\\
QGSM &6212 &77 $\pm$ 2&80 $\pm$ 3
&68 $\pm$ 5&69 $\pm$ 5 &82 $\pm$ 4 &89 $\pm$ 3 \\
 &&&&&&&\\
\hline
\end{tabular}
\newpage
{\bf{Table 2.}} The results of the fit of the  distributions of $\pi^{-}$
mesons in the region  $\zeta^{+} < \tilde{\zeta^{+}}$ .
\   \par
\   \par
\  \par
\  \par
\begin{tabular}{|c|c|c|c|c|}    \hline
\multicolumn{1}{|c|}{}&
\multicolumn{3}{|c|}{}&
\multicolumn{1}{c|}{}\\
\multicolumn{1}{|c|}{}&
\multicolumn{3}{|c|}{{$dN/dp_{T}^{2}$}}&
\multicolumn{1}{c|}{{$1/\pi \ast dN/d\zeta ^{+}$ }}\\
\multicolumn{1}{|c|}{}&
\multicolumn{3}{|c|}{}&
\multicolumn{1}{c|}{}\\
\cline{2-5}
   &$  \alpha $ & $\beta_{1}$ &$ \beta_{2}$  & $n$ \\
  &  &$ (GeV/c)^{-2}$ & $(GeV/c)^{-2}$ & \\
\hline
 & & & & \\
       exp. &  0.85 $\pm$ 0.03 & 12.0 $\pm$ 0.4 & 4.8 $\pm$ 0.3 &4.30 $\pm$ 0.06 \\
& & & & \\
\hline
 & & & & \\
QGSM&0.90 $\pm$ 0.05 & 11.4 $\pm$ 0.6 & 5.2 $\pm$ 0.9 &4.17 $\pm$ 0.11  \\
 & & & & \\
\hline
\end{tabular}
\newpage
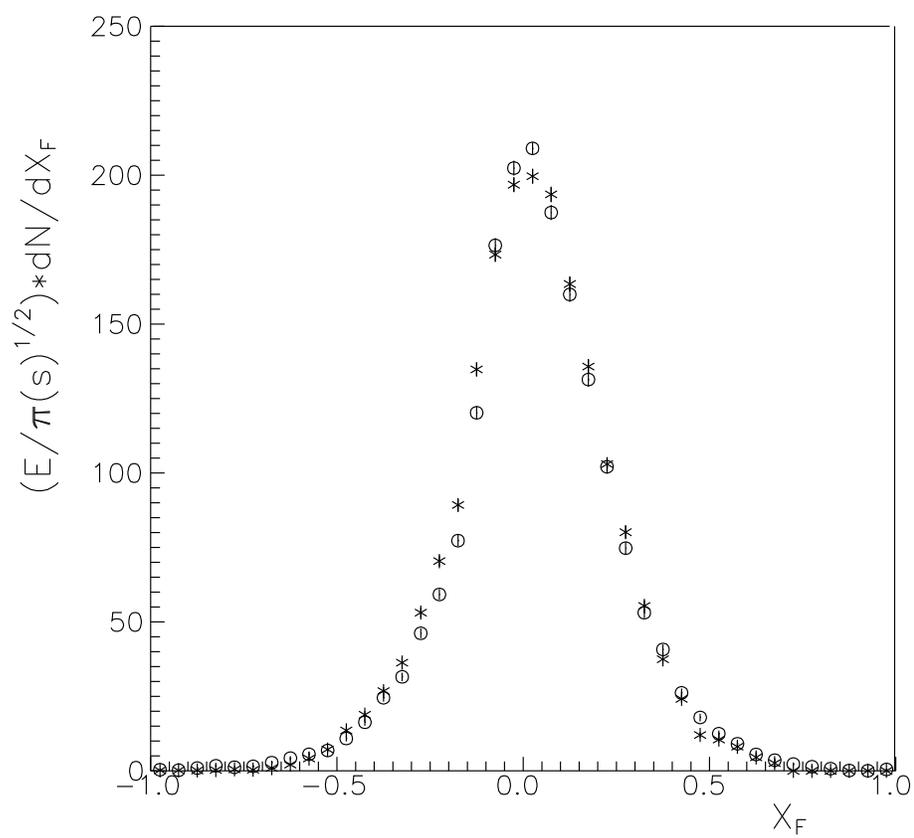
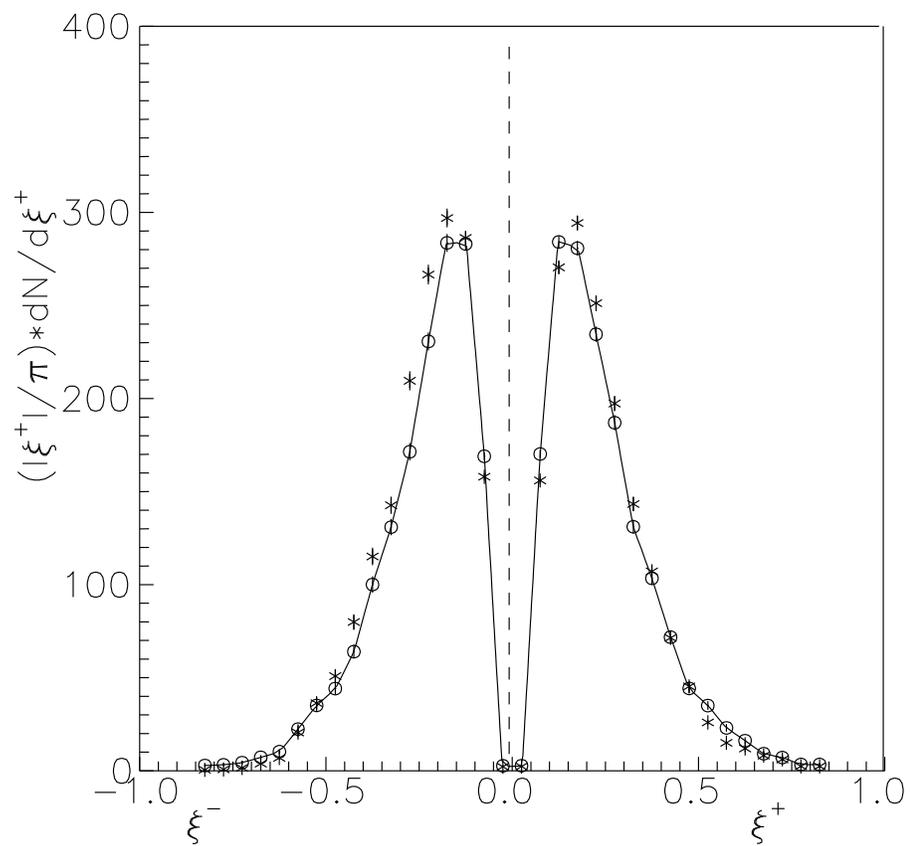
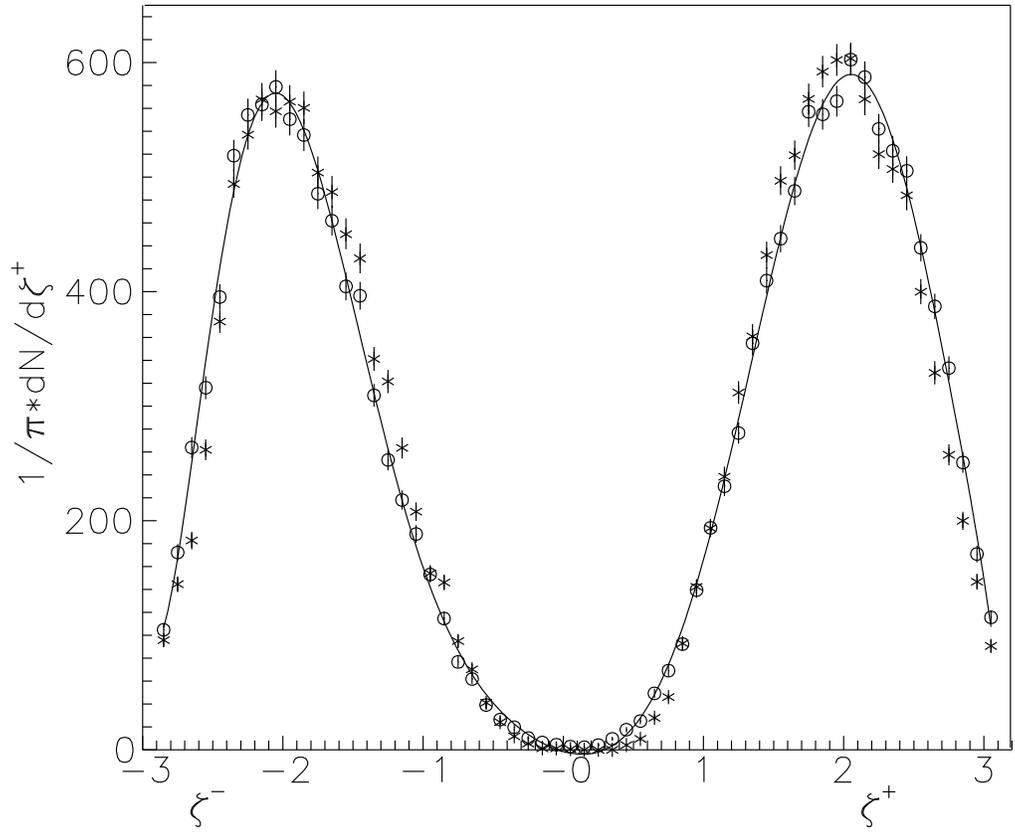
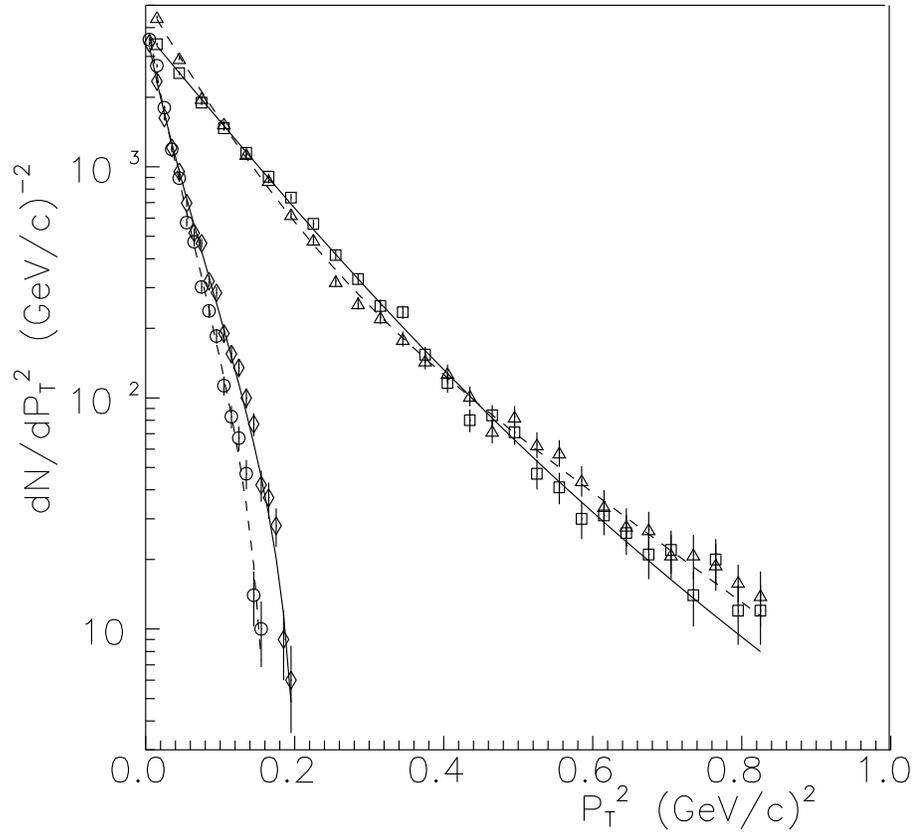
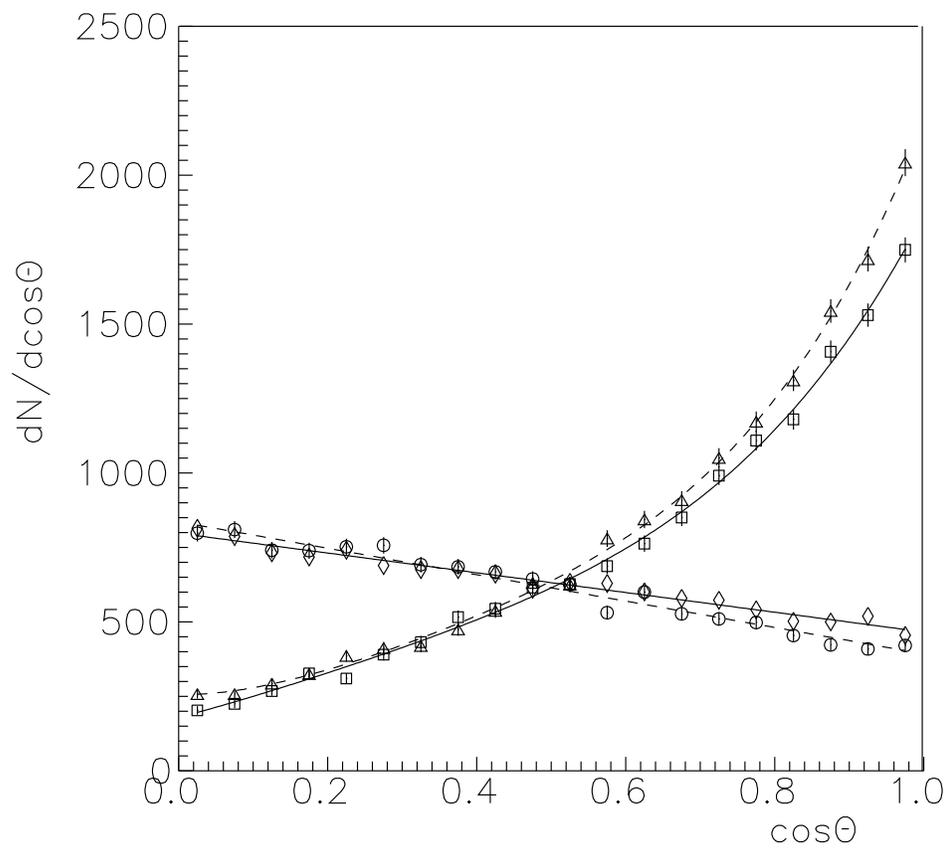
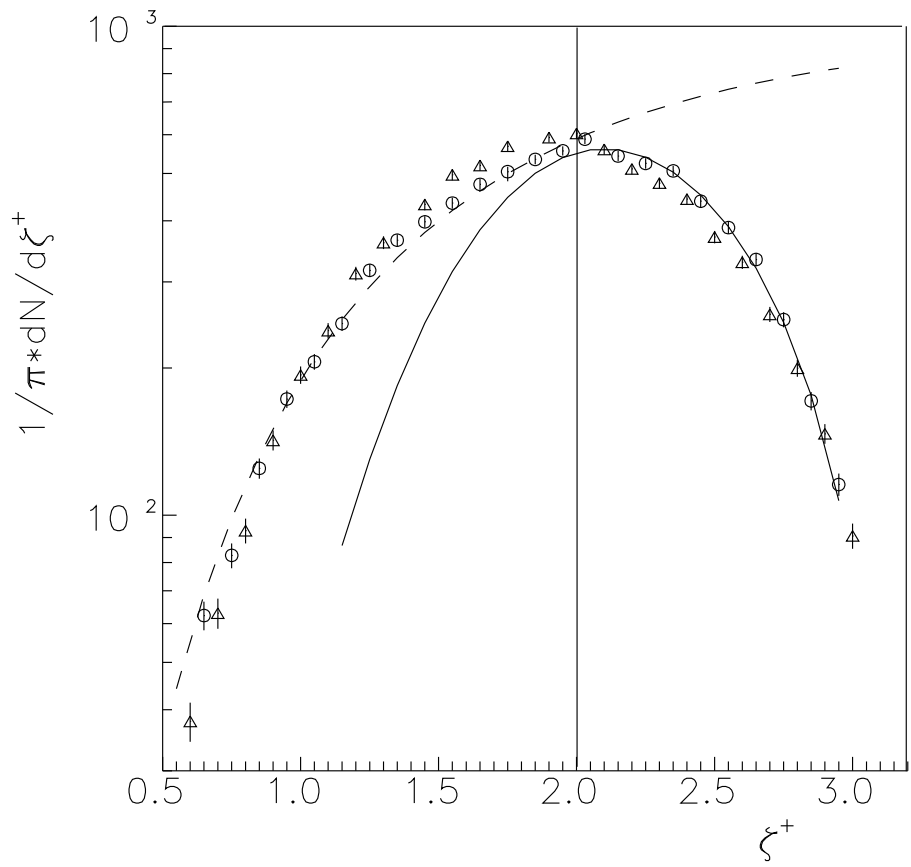
\end{document}